\documentclass[12pt,preprint]{aastex}
\usepackage[]{natbib}

\begin{document}

\title{PROPOSED GRAVITATIONAL WAVE BACKGROUND FROM BLACK
HOLE--TORUS SYSTEMS}

\author{David M. Coward\altaffilmark{1}, Maurice H.P.M. van Putten\altaffilmark{2}, 
and Ronald R. Burman\altaffilmark{1}}
\affil{\altaffilmark{1}Department of Physics, University of
Western Australia, Nedlands WA 6009, Australia}
\affil{\altaffilmark{2}Massachusetts Institute of Technology, Room 2-378, MA 02139}

\date{\today}% It is always \today, today,

\begin{abstract}
Cosmological gamma-ray bursts may be powered by rotating black holes with 
contemporaneous emission of gravitational radiation from a surrounding torus. 
We calculate the resulting stochastic background radiation assuming 
strong cosmological evolution and a uniform black hole mass distribution of 
$M=$ (4--14)M$_\odot$. The predicted spectral flux density corresponds to a peak 
spectral closure density of (1--2$)\times 10^{-7}$, and has comparable
contributions at 450 Hz$\times\kappa$ and over 300--450
Hz$\times\kappa$ from nearby and distant sources, respectively,
where $\kappa$ refers to an uncertainty factor of order unity in the radius of the
torus. For two optimized advanced LIGO-type detectors the proposed
gravitational wave background could be detectable within a year of integration.
\end{abstract}

\section{INTRODUCTION}
Cosmological gamma-ray bursts are the most enigmatic transient
events in the Universe. They show a bi-modal distribution in
durations of short bursts around 0.3 s and long bursts around 30 s
\citep{kou93}. Based on their GRB fluence and time-variability,
their inner engines should be compact and highly energetic.
Leading candidates for GRB progenitors are collapsars and mergers
of black holes and neutron stars. In particular, long bursts have
been associated with core collapse of massive stars
\citep{woos93,Mwoos99}, or their hypernova variants, in
star-forming regions \citep{Paczy97,Brown2000}. These, and mergers
of compact binaries, are believed to result in black hole plus
disk or torus systems --- see van Putten (2001) for a review.

A torus around a rapidly rotating black hole converts spin energy into
various channels, notably gravitational and thermal radiation, winds and
MeV neutrino emissions \citep{mvp02}. These emissions 
last for the life-time of rapid spin of the black hole -- the
de-redshifted durations of tens of seconds for long bursts from
black hole-torus systems in suspended accretion \citep{mvpostr}.
The single-source spectrum is here described
by a horizontal branch in the $\dot{f}(f)$ diagram \citep{mvp00}
within a frequency range in the vicinity of 1 kHz for a
7-M$_{\odot}$ black hole. GRB energies appear to have a diversity
of about one order of magnitude \citep{Frail2001,Piran2001}. Black
hole-torus systems are expected to have a distribution in black
hole mass $M$ consistent with the recently proposed association
with soft X-ray transients in the hypernova proposal of
\citet{Brown2000}; i.e., $M\simeq$ (4--14)M$_{\odot}$.

In this work, we calculate the gravitational wave (GW) spectra
expected from a cosmological distribution of black hole-torus
systems, assuming strong cosmological evolution locked to the
star-formation rate (SFR). We shall estimate the expected
contribution to the stochastic background in GWs from the low- and
high-redshift populations. Following \citet{scm01} and
\citet{Frail2001}, the total event rate is normalized to a local
GRB rate of 0.5 yr$^{-1}$ Gpc$^{-3}$ at $z=0$, assuming a
``flat-$\Lambda$" cosmology.

\section{THE EVOLVING GRB RATE}
The SFR can be parametrized according to the model SF2 of
\citet{M2001}, wherein the density rises rapidly by an order of
magnitude between $z=0$ and $z=1$, peaks between $z=1$ and 2 and
declines gently at higher redshifts:
\begin{equation}
%\ R_\mathrm{SF2}(z) =
\ R_{\mathrm{SF2}}(z) =
0.15h_{65}[1+22\mathrm{exp}(-3.4z)]^{-1}\hspace{.2cm}{\mathrm
{M_\odot\hspace{.05cm} yr^{-1}\hspace{.05cm} Mpc^{-3}}}
\end{equation}
with $h_{65}$ denoting the Hubble constant normalized to 65 km
s$^{-1}$ Mpc$^{-1}$. This model was derived using the Einstein--de
Sitter cosmology, so a scaling factor has to be applied for other
cosmologies \citep{M2001}. We define a dimensionless evolution
factor $e(z)$ as $R_{\mathrm{SF2}}(z)$/$R_{\mathrm{SF2}}(0)$ and
assume a cutoff for star formation at $z$ = 5, because active star
formation is believed to have begun at that epoch. The variation
of the rate of GRB sources with redshift can be expressed by the
event-rate equation
\begin{equation}%2
dR/dz = 4\pi (c^3r{_0}/H_0{^3})e(z)F(z)/(1+z) \label{EQ drdz}
\end{equation}
\noindent where $r_{0}$ is their present-epoch rate density and
$R(z)$ is the all-sky event rate, as observed in our local frame,
for sources out to redshift $z$. The factor $c^{3}r_{0}/H_{0}^{3}$
has the dimensions of inverse time, as does the event rate $R(z)$.
The $(1 + z)$ denominator in (2) accounts for the time dilation of
the observed rate by cosmic expansion, converting a source-count
equation to an event-rate equation. The dimensionless function
$F(z)$ is determined by the cosmological model and can be
calculated from the Hubble parameter $H(z)$ and the angular size
distance \citep[p.~332]{Pe}.

Fig. 1 plots the resulting GRB rate evolution for three standard
cosmologies: ($\Omega_{\mathrm{M}},\Omega_{\Lambda}$) $=$ (1, 0),
(0.3, 0.7) and (0.3, 0), all with $h_{65}=1$, using local GRB rate
densities of 0.9, 0.5 and 0.6 yr$^{-1}$ Gpc$^{-3}$ respectively;
the differing local rate densities reflect the different spatial
geometries in the three cosmologies. The steep rise at low $z$
reflects a combined increase in sky surface area and
star-formation rate. Recent observation-based analysis of GRB jet
opening angles indicates a beaming factor of about 500
\citep{Frail2001,Piran2001}. This factor increases the
above-stated local rates, whereby the corresponding cumulative
all-sky rate becomes a few per minute (several thousand per day)
after integration of equation (2) out to $z=5$. We adopt a
present-epoch GRB rate density of 250 yr$^{-1}$ Gpc$^{-3}$ in the
flat-$\Lambda$ cosmology with $h_{65}=1$.

\section{SINGLE-SOURCE SPECTRA: A GENTLE CHIRP}
%\noindent{\it Single-source Spectra: a Gentle Chirp.}
The single-source spectrum of gravitational waves of a black
hole-torus system is modeled by the emission of a fraction of
about ten percent of the rotational energy of a rapidly spinning 
black hole \citep{mvp02}. These emissions correspond to mass 
inhomogeneities in the torus on the order of a few permille of the
black hole mass, which have recently been identified
as Papaloizou-Pringle buckling modes \citep{mvp02b}. The lowest
frequency emission relevant to the current laser-interferometric
gravitational wave experiments is produced by a quadrupole moment 
of the torus. This frequency equals twice the Keplerian angular 
velocity of the torus, in this paper normalized to $f=1$~kHz for a 
7-M$_{\odot}$ black hole. A spread as well as a systematic uncertainty 
presently exists in the radius of the torus, and hence in the expected 
frequency of gravitational radiation. Because most of the rotational energy is 
concentrated at high spin-rate, the frequency evolves according to an 
approximately linear chirp---i.e., a ``gentle chirp" in the form of a
quasi-periodic oscillation with a bandwidth of about 10\% within
the de-redshifted duration of mean value 20 s. This is a
horizontal branch in the $\dot{f}(f) $ diagram \citep{mvp00}:
\begin{equation}
f \approx f_i + \dot{f}t \label{EQN_GW}
\end{equation}
for an initial frequency $f_i$ and an approximately constant time
rate of change $\dot{f}$. The sign of $\dot{f}$ is not
predicted by the model. In what follows, $\dot{f}<0$ is assumed; the
final results are not expected to change qualitatively for the
opposite choice of sign in the frequency sweep.

A scaling for the chirp with the black hole mass obtains, assuming
a constant ratio ${\cal E}_k/{\cal E}_B\sim100$ of the kinetic
energy to poloidal magnetic energy in the torus \citep{mvpostr}.
This energy ratio is a major uncertain parameter in accretion disk
physics. It holds constant if a disk dynamo process operates which
establishes an equilibrium with the driving kinetic energy in
response to the powerful competing torques acting on the inner and
outer faces of the torus, with no memory of initial conditions.
The inner face of the torus receives most of the black hole
luminosity $L_H\propto{\cal E}_B/M$ for a net energy ${\cal E}_B$
in the magnetic field, by equivalence in poloidal topology to
pulsar magnetospheres \citep{mvp99}. As the rotational energy of a
rapidly rotating black hole is proportional to its mass, its
life-time of rapid spin prolongs in proportion to $M^{2}$,
assuming the torus radius to scale linearly with $M$. In the
present context, it would be sufficient for this scaling to hold
as an ensemble average. This condition is equivalent to the
assumption of constant fractions of magnetic flux connected to the
horizon of the black hole and in winds to infinity.

The energy in gravitational radiation consists of the
above-mentioned fraction of the spin-energy of the black hole, and
is proportional to the ratio of torus-to-black hole angular
velocities. At constant ratio of the torus radius $R$ to the hole
mass $M$, the energy output in gravitational radiation is
proportional to the latter; this gives
\begin{equation}
f_i\propto M^{-1},\hspace{.2cm}\dot{f}\propto
M^{-3},\hspace{.2cm}E_{GW} \simeq 0.03 M c^2.\label{EQN_GW2},
\end{equation}
%\indent
where $f_i=$ 1 kHz for a 7-M$_{\odot}$ black hole and $R=4M$, with durations 
consistent with the de-redshifted mean duration of long GRB events of about 
20 s. The frequency $f$ may vary over the life-time of rapid spin of the 
black hole, by an amount which could reach about 10\%. In what follows, a 
frequency sweep of about 100 Hz is assumed, so $\dot{f}= -5$ Hz s$^{-1}$,
for $M=7$M$_{\odot}$.

The single-source spectral time-integrated flux density (or
``spectral fluence"), in J m$^{ - 2}$ Hz$^{ - 1}$, of a quadrupole
GW signal at a luminosity distance $d_{L}(z)$ is expressed as
\citep{FSM99a,FSM99b,Coward2001}
\begin{equation}
F_{\mathrm{ss}} (f_{\mathrm{obs}},z) = (c^3/8\pi
G)f_{\mathrm{obs}}^2{\left| {\tilde {h}(f_{\mathrm{obs}})}
\right|}^2{ \mathord{\left/ {\vphantom { {d_{L} ^{2}(z)}}} \right.
\kern-\nulldelimiterspace} {d_{L}^2(z)}}.\label{EQ
Fss}\end{equation}
\noindent Here $\tilde{h}(f_{\mathrm{obs}})$,
which is in m/Hz, is the Fourier transform of the wave amplitude
(in meters) at the observed frequency $f_{\mathrm{obs}}$, which is
related to the source frequency $f$ by the redshift factor:
$f_{\mathrm{obs}}=f/(1+z)$. The factor $(c^{3}/8\pi G)$ has
dimensions mass$\slash$time, equivalent to energy/(area $\times$
frequency).

Fig. 2 illustrates the spectral fluence for a selection of black
hole masses ranging from 14M$_\odot$ to 4M$_\odot$ in steps of
1M$_\odot$, for negative $\dot{f}$ at a source distance of 100 Mpc
$(z\approx0.02)$; via (4), the corresponding $f_{i}$ range over
0.5--1.75 kHz with $-\dot{f}$ ranging over 0.625--27 Hz s$^{-1}$;
the bandwidth $\Delta f$ has been taken to be 10\% of $f_{i}$ in
each case, ranging over 175--50 Hz. The figure depicts the
decrease of both $f_{i}$ and $\Delta f$, and also the steepening
of the spectrum, with increasing $M$.

The scaling of the burst duration $T$ follows from (4):
\begin{equation}
T\approx\Delta f/\dot{f}\approx0.1f_{i}/\dot{f}\propto M^{2},
\label{EQ Fss}\end{equation} \noindent ranging over 44--3.6 s for
$M=($14--4)M$_{\odot}$; we have chosen $T=11$ s for $M=$
7M$_{\odot}$, so that $T$ averages to 20 s over the uniform
distribution of $M$. The scaling of $N$, the total number of
cycles comprising the burst, also follows from (4): for a 10\%
frequency sweep
\begin{equation}
N\approx 0.95f_{i}T\propto M, \label{EQ Fss}\end{equation} with
$N=1.0\times10^{4}$ for $M=7$M$_{\odot}$.

A single-source dimensionless characteristic amplitude is
expressed by \citep{flan98}
\begin{equation}
h_{\mathrm{char}}(f_{\mathrm{obs}},z)\equiv\sqrt{\frac{2G}{c^{3}}}\frac{1+z}{\pi
d_{L}(z)}\sqrt{\frac{dE(f_{\mathrm{obs}})}{df_{\mathrm{obs}}}}\hspace{0.1cm},
\label{EQN_H}
\end{equation}
where $dE(f_{\mathrm{obs}})/df_{\mathrm{obs}}$ is the energy
spectrum, representing the distribution of GW energy per unit of
observed frequency at luminosity distance $d_{L}(z)$. Fig. 3 shows
$h_{\mathrm{char}}(f_{\mathrm{obs}},z)$ for the same sources as in
Fig. 2, at the same distance. The top curve assumes optimal
matched filtering over the duration of the emission. The bottom
curve is $h_{\mathrm{char}}(f_{\mathrm{obs}},z)$ for a single
cycle of the above emission. Matched filtering will increase the
signal-to-noise ratio by a factor $\sqrt{fT}\approx\sqrt{N}$. Fig.
3 shows, for example, that the dimensionless strain amplitude of a
single burst event is expressed by $h_{\mathrm{char}}\simeq
6\times 10^{-21}$ for $M=7$M$_\odot$ with $f_{\mathrm{obs}}=$
980--880 Hz; for an 11-s emission at a mean frequency of 950 Hz,
$h_{\mathrm{char}}\slash\sqrt{N}\approx0.06\times10^{-21}$ for a
single cycle. For comparison with these predictions, a model for
the root-mean-square dimensionless noise amplitude of an advanced
LIGO detector \citep{flan98} is included in the figure.

\section{GW BACKGROUND SPECTRA}
%\noindent{\it GW Background Spectra.}
The stochastic background produced by the proposed radiation from
black hole-torus systems may be expressed by their spectral flux
density, in W m$^{ - 2}$ Hz$^{ - 1}$. The contribution from these
sources throughout the Universe obtains by integrating the product
$F_{\mathrm{ss}}( f_{\mathrm{obs}}, z) dR/dz$ over the redshift
range $z = 0$ to $z = $ 5:
\begin{equation}
F_{B} (f_{\mathrm{obs}}) = {\int\limits_{0}^{5} F_{\mathrm{ss}}
(f_{\mathrm{obs}}},z)({{dR} \mathord{\left/ {\vphantom {{dR}
{dz}}} \right. \kern-\nulldelimiterspace} {dz})\mbox{} dz},
\end{equation}
\noindent with $F_{\mathrm{ss}}$ and $dR/dz$ given by (\ref{EQ
Fss}) and (\ref{EQ drdz}). The background spectral strain, in
Hz$^{ - 1 / 2}$, is calculated directly from this background
spectral flux density (Ferrari et al. 1999a,b):%\citep{FSM99a,FSM99b}
\begin{equation}
\sqrt {S_{B} (f_{\mathrm{obs}}} ) = (2G/\pi c^{3}
)^{1/2}f_{\mathrm{obs}}^{-1} [F_{B}(f_{\mathrm{obs}})]^{1/2}.
\end{equation}

The spectral energy density of a GW background is conventionally
expressed by the dimensionless spectral ``closure density",
defined as the energy density of gravitational waves per
logarithmic frequency interval normalized to the cosmological
critical density $\rho_{c}c^{2}$. It can be obtained from the
background spectral
flux density (Ferrari et al. 1999a,b) as %\citep{FSM99a,FSM99b}
\begin{equation}
\Omega_{B}(f_{\mathrm{obs}}) = f_{\mathrm{obs}}
F_{B}(f_{\mathrm{obs}})/(\rho_{c}c^{3}).
\end{equation}

The duty cycle ($DC$) from events out to redshift $z$ is given by
\begin{equation}
DC(z) = \int\limits_{0}^{z}(1+z) \tau (dR/dz) dz, \label{EQN dc}
\end{equation}
\noindent where the typical duration, $\tau$, of the signal is
dilated to $(1+z)\tau$ by the cosmic expansion. A calculated $DC$
of unity or greater implies that the signal is continuous. If
$DC\gg 1$, then the central limit theorem tells us that the
amplitude distribution can be approximated as a Gaussian. If the
calculated $DC\leq  1$, then the amplitude distribution can be
simulated (Coward et al. 2002a,b) from a random sampling of the
probability distribution based on the event rate equation
(\ref{EQ drdz}).%\citep*{Coward2002a,Coward2002b}

\section{NUMERICAL RESULTS}
%\noindent{\it Numerical results.}
The stochastic background produced by  systems with a uniform
black hole mass distribution $M=$ (4--14)M$_\odot$ for three
standard cosmologies, expressed in spectral flux density, is shown
in Fig. 4. For the flat-$\Lambda$ cosmology, it exhibits a broad
plateau of $5 \times 10^{-11}$ W m$^{-2}$ Hz$^{-1}$ at about
300--450 Hz. At 300 Hz, the main contributions are from
cosmologically distant sources at about $z=$ 0.6--1. The edge at
450 Hz represents contributions from more local sources and is
strongly dependent on the single-source fluence. For the EdS
cosmology, the plateau is similar in bandwidth but scaled up to $1
\times 10^{-10}$ W m$^{-2}$ Hz$^{-1}$. The open cosmology shows a
noticeable peak corresponding to contributions from local
higher-mass black holes.

The variation in spectral flux density among the flat-$\Lambda$,
EdS and open cosmologies can be attributed to the cosmology
dependence of the luminosity distance. Fig. 5 plots the ratio of
inverse luminosity distance squared for the open and EdS to
flat-$\Lambda$ cosmologies. It shows a maximum of 1.25 at
$z\approx 1$ for the open to flat-$\Lambda$ case, consistent with
the ratio of the spectral flux density in Fig. 4 for these
cosmologies. For the EdS to flat-$\Lambda$ case, the ratio in Fig.
5 increases steadily to 2.35 at $z=5$, which is consistent with
the ratio of the plateau heights in Fig. 4 for these cosmologies.

The spectral strain, shown in Fig. 6 for the same cosmologies,
exhibits a maximum of (3--5$)\times10^{-26}$ Hz$^{-1/2}$ at about
200--250 Hz, and a steep rise at a frequency of about 100 Hz
caused by the cutoff at $z$ = 5 of the SFR. The SFR may in fact
decrease smoothly beyond a redshift of about 5, hence moderating
this rather non-physical feature in the spectral strain.
Incorporating higher-$z$ sources is not expected to affect our
results significantly, as $z > 5$ sources are likely to be fewer
in number than sources at $z$ = 1--2 (see Fig. 1) and the
increasing luminosity distance of high-$z$ sources means that
their contribution to the background flux density becomes less
significant. The tails at frequencies above 500 Hz in Figs. 4 and
6 represent the contributions from low-mass black holes;
evidently, the spectra are dominated by high-mass black holes.

The spectral closure density of this background radiation, shown
in Fig. 7, has a sharp maximum of (1--2$)\times10^{-7}$ at about
450 Hz for the same cosmologies. The broad hump around 300 Hz,
visible in the curve for the flat-$\Lambda$ cosmology, and the
sharp peak at about 450 Hz are representative of cosmological and
local sources respectively.

The duty cycle of the burst events throughout the Universe reaches
about 2, upon integration of (12) up to $z=5$. Fig. 8 plots
$DC(z)$, exhibiting very little variation in form among the three
cosmologies. The rate estimate 250 yr$^{-1}$ Gpc$^{-3}$ at $z=0$
via Schmidt (2001) and Frail (2001) translates to a net GRB event
rate throughout the Universe of a few per minute, as seen in our
frame. This signal is comprised of discrete pulses, originating
largely from sources at redshifts of 1--3. Since these pulses last
typically about 40--80 s (in our frame), they overlap and produce
a near-continuous background, to within the approximations made in
our treatment.

\section{Detectability}

A stochastic background will manifest itself in a single detector
as excess noise. The signal from the proposed GRB GW background is
expected to be far below the noise level in any single planned
ground-based detector. The signal-to-noise ratio $(S/N)$ in terms
of amplitude for either an interferometer or resonant-mass
detector is \citep{Magg2000}:
\begin{equation}
S/N=\left[FS_{h}(f)\slash S_{n}(f)\right]^{1/2};
\end{equation}
\noindent here, $F$ is a pattern function, which is a measure of
the angular efficiency of the detector, $S_{h}(f)$ is the signal
power spectral density (density in frequency space) and $S_{n}(f)$
is the noise power spectral density, both in units of Hz$^{-1}$.
Fig. 9 is a plot of the $S/N$ of the proposed background as a
function of frequency, using $S_{h}(f)$ calculated from the
flat-$\Lambda$ spectral strain shown in Fig. 6 and a model for the
noise power spectrum an ``advanced LIGO" type interferometer
(Flanagan \& Hughes 1998). The $S/N$ is small, with a sharp
maximum of $1 \times 10^{-2}$ near 100 Hz; the rapid fall to the
left of this is due mainly to the absence of signal below about
100 Hz.

The most promising detection strategy is that of cross-correlation
of the output of two neighboring detectors, as recently reviewed
by Allen \& Romano (1999) and Maggiore (2000). For the signals in
them to be correlated, the detectors must be separated by less
than one reduced wavelength, which is about 100 km for frequencies
around 500 Hz where we expect the GRB-generated background to
peak. The detectors also need to be sufficiently well separated
that their noise sources are largely uncorrelated.

Under these conditions, assuming Gaussian noise in each detector
and optimal filtering, a filter function chosen to maximize $S/N$
for two such detectors yields the formula
\citep[Eq.~3.75]{Allen1999}:
\begin{equation}%10
\left(\frac{S}{N}\right)^{2}\approx\frac{9H_{0}^{4}} {50 \pi^{4}}
T \int_{0}^{\infty} \frac{\gamma^{2}(f)\Omega_{B}^{2}(f)}{f^{6}
S_{n1}(f)S_{n2}(f)}.
\end{equation}
Here, $\gamma (f)$ is an ``overlap reduction function'', which
accounts for the separation and relative orientation of the
detectors, and $S_{n1}(f)$ and $S_{n2}(f)$ are the noise power
spectral densities of the detectors. As the optimal filter depends
on $\Omega_{B}(f)$, a range of filter functions based on
theoretical expectations of this function will need to be used.

For this preliminary study of the detectability of the
GRB-generated background, we assume an optimized value of close to
unity for $\gamma (f)$ of two detectors situated within several
kilometres. We take the flat-$\Lambda$ $\Omega_{B}(f)$ as shown in
Fig. 7 and use a piecewise parametrized model for $S_{n1}(f)$ and
$S_{n2}(f)$ for proposed advanced LIGO detectors (Flanagan \&
Hughes 1998), assuming a pair of similar detectors. Fig. 10 plots
the resulting $S/N$ as a function of integration time, yielding a
value of about 8 for 1 yr of integration. This preliminary result
suggests that the proposed GRB GW background is potentially
detectable given optimized cross-correlation between two advanced
LIGO detectors.

We also note that given the uncertainty in the sign of
$\dot{f}(f)$ in the single-source emission model, the peak flux
may occur at higher frequencies. If so then the the GRB GW
background may be detectable using cross-correlation with a
resonant-mass interferometer detector pair. The LIGO-WA
interferometer and ALLEGRO resonant-mass detector pair, when
co-aligned, yield a value for $\gamma$ close to unity across the
bandwidth 1--1000 Hz \citep{stochsearch1}. Unfortunately this
bandwidth is not fully utilized in the ratio of GW power to noise
power appearing in the integral in Eq. (14), because ALLEGRO, as a
resonant-mass detector, has a narrow bandwidth, centered on a
resonant frequency of about 900 Hz in this case.

\section{DISCUSSION}
%\noindent{\it Discussion.}
\noindent We draw several conclusions:

\noindent 1. Black hole-torus systems associated with GRBs are
expected to give a substantial contribution to the stochastic GW
background in a frequency window of 300-450 Hz$\times\kappa$,
where $\kappa$ denotes an uncertainty factor of order unity in 
the radius of the torus. If black hole-torus systems exist also as 
transient sources independent of GRBs, their event rate will be
higher with a commensurably more pronounced stochastic GW background.

\noindent 2. It is instructive to compare the presented results
with radiation from rapidly rotating neutron stars. 
Black hole-torus systems produce spectral flux densities 
which are similar to the contribution from r-mode instabilities 
as described in Ferrari et al. (1999b). Here, the large
output provided by the spin-energy of the black hole
compensates for an event rate which is less than the formation
rate of neutron stars by some three orders of magnitude.  
The frequencies and $DC$ of the former are markedly higher,
respectively, lower than those predicted for radiation from
neutron star modes, however. The $DC$ of the signals from neutron
star r-modes may be as high as $10^{9}$ (Ferrari et al. 1999b) in
contrast to our $DC$ estimate of order unity for the signals from
GRB sources.
This comparision becomes more favorable with recent understanding 
of r-modes in more detailed and realistic scenarios.
The r-modes are driven by a generally weak gravitational 
radiation-reaction force and, in the perturbative limit, effectively 
decoupled from other modes \citep{sch02}. A recent appreciation of 
various channels for creating viscosity \citep{owen02,rez01,wu01} 
renders their saturation energies small, and less so than previously 
thought \citep{arra02}. 

\noindent 3. The high output in gravitational radiation powered by
the spin-energy of the black hole gives rise to a major
contribution to the predicted GW background by a relatively nearby
population of sources. This is apparent in the edge at 450
Hz$\times\kappa$ in Fig. 4 to the 300--450 Hz$\times\kappa$
flat-$\Lambda$ spectral flux density plateau due to sources at
higher redshifts. The comparable contributions by the nearby and
distant sources is a robust result, depending only on the
assumption that the proposed black hole-torus systems are tightly
locked to the SFR.

\noindent 4. As the local GRB rate (assumed in this work) is about
1 yr$^{-1}$ within a radius of 100 Mpc, the average background
spectral flux density should be representative of sources from
$z=$ 0.02--5 for an observation time of one year. One can view the
background spectral flux density shown in Fig. 4 as the result of
averaging the flux from a discrete number distribution over an
infinite observation time. We plan to investigate these issues
further in connection with the detectability of this predicted
background using advanced LIGO-type detectors.

\noindent 5. The estimated frequency range of about 200--2000 Hz
of the proposed background and foreground radiation in
gravitational waves from black hole-torus systems in association
with GRBs defines a new source with a well-defined event rate in
the high-frequency bandwidth of LIGO/VIRGO detectors as well as
for some of the current bar detectors. Cross-correlating the
output between pairs of detectors promises to be the optimum
detection strategy. For an idealized advanced LIGO pair of
detectors, the $S/N$ could be high enough for detection of the
proposed background in a year of integration. Using present
detector technology and locations, the optimum detector pair
combination may be a resonant-mass and interferometric detector,
such as the LIGO-WA and ALLEGRO pair.

\acknowledgments The authors thank the referee for a helpful 
review. This research was funded in part by the
Australian Research Council and is part of the coordinated
research program of the Australian Consortium for Interferometric
Gravitational Astronomy. MVP acknowledges support by NASA Grant
5-7012 and the MIT C.E. Reed Fund, and stimulating discussions
with R. Remillard.

%\begin{references}

\begin{figure*}
\plotone{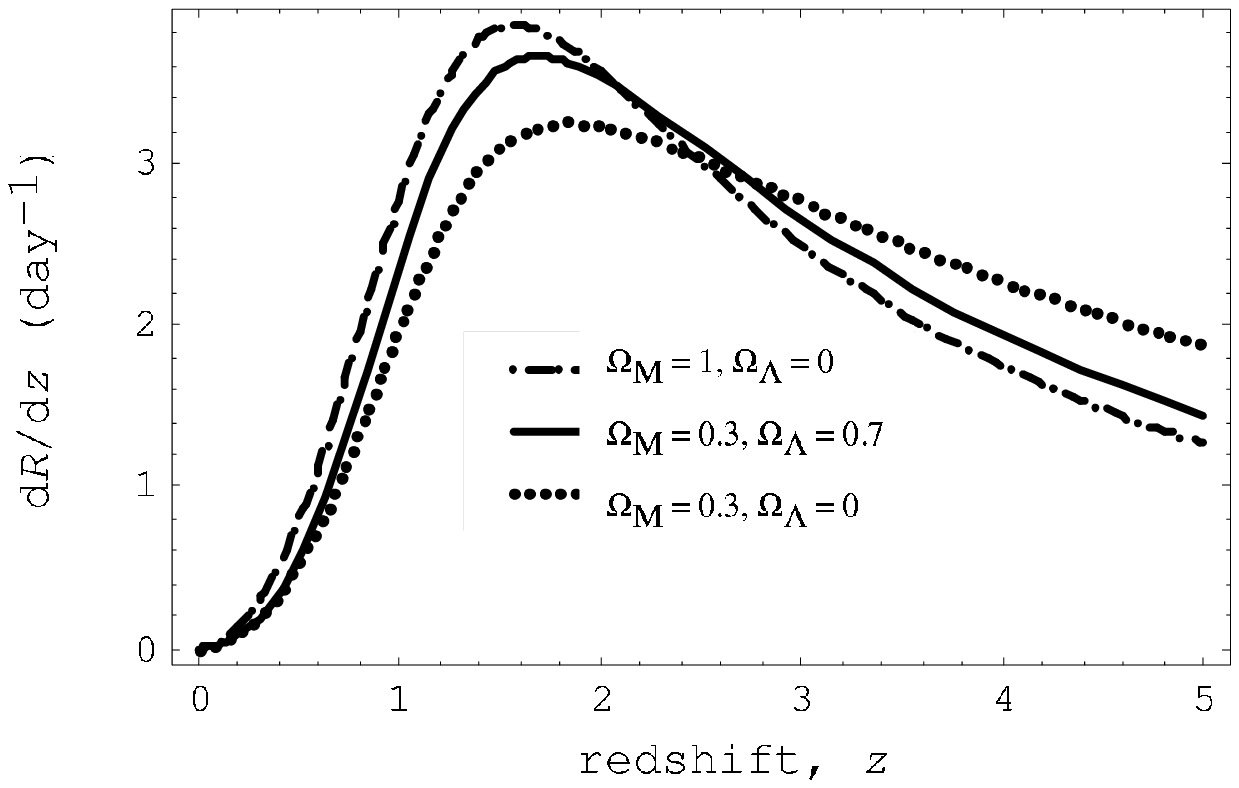}%
\caption{\label{Fig. 1}The differential event rate $dR/dz$ of GRBs
as a function of redshift for three standard cosmologies using the
SFR model SF2 of Porciani \& Madau (2001), normalized to a local
GRB rate of 0.5 yr$^{-1}$ Gpc$^{-3}$ in the flat-$\Lambda$
cosmology. The mean redshifts (as seen in our frame) are 2.46,
2.47 and 2.72 in, respectively, the EdS, flat-$\Lambda$ and open
cosmologies. The cumulative all-sky rate of GRBs reaches a few per
minute, upon integration of $dR/dz$ up to a redshift $z=5$ and
multiplying by a beaming factor of 500.}
%\label{fig1}
\end{figure*}

\begin{figure*}
\plotone{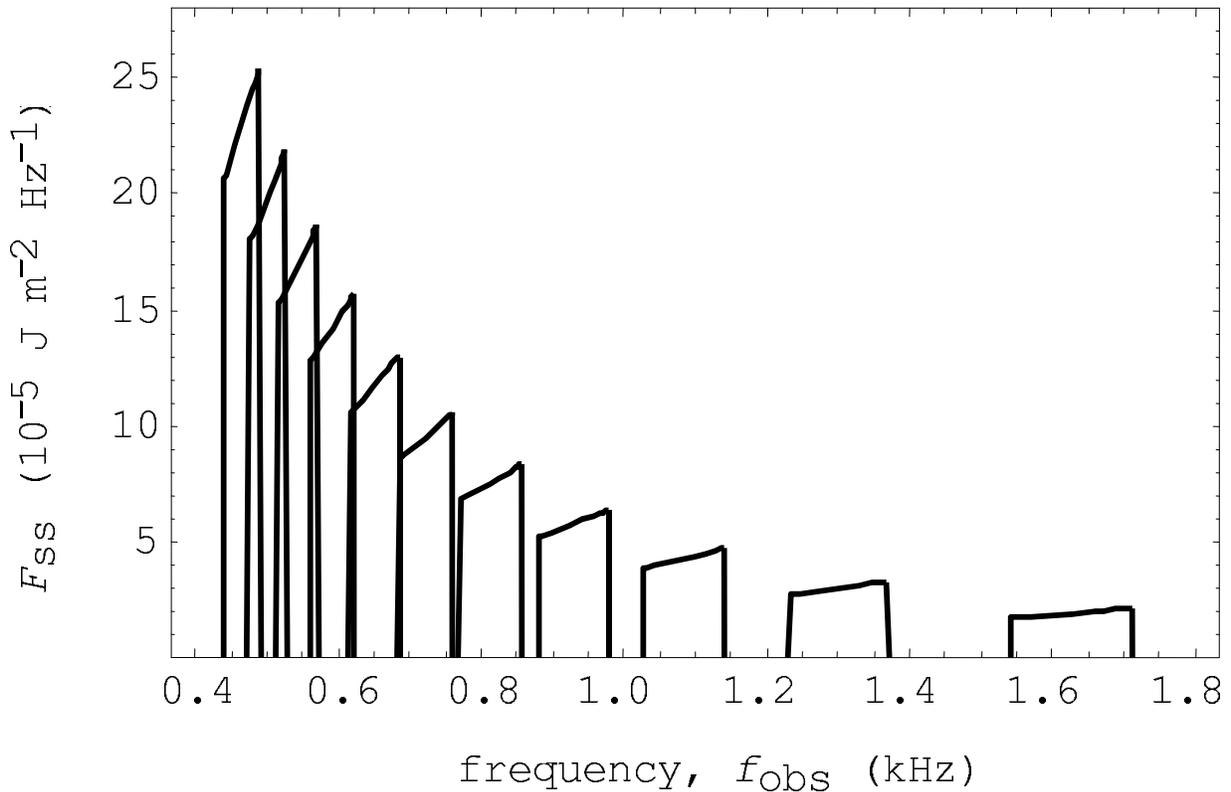}%
\caption{The single-source GW spectral fluence of black hole-torus
systems for a selection of black hole masses, ranging from
14M$_\odot$ (far left) to 4M$_\odot$ (far right) in steps of
1M$_\odot$, for negative $\dot{f}$. The source distance is 100 Mpc
($z\approx 0.02$). Each source produces a gentle chirp on a
horizontal branch in the $\dot{f}(f)$ diagram. The frequencies
shown vary by a factor $\kappa$ of order unity due to an
uncertainty in the radius of the torus.} \label{fig2}
\end{figure*}

\begin{figure*}
\plotone{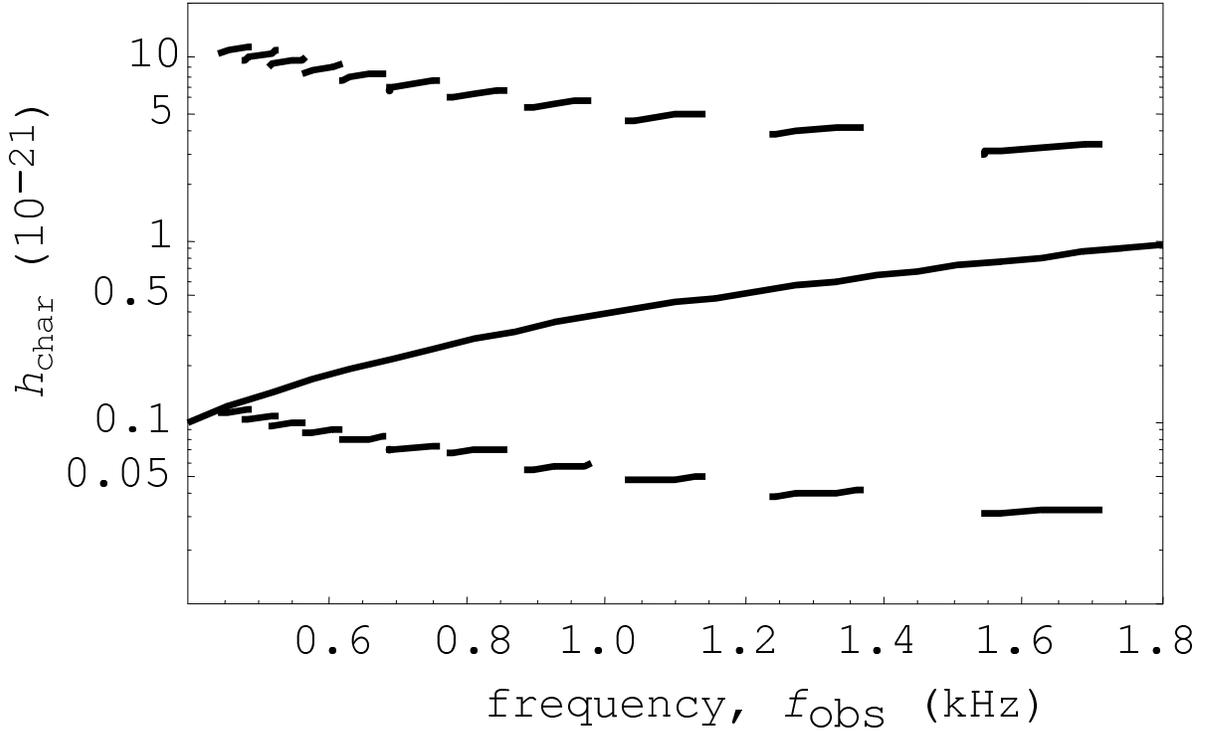}%
\caption{{\bf Top curve}---the characteristic dimensionless
amplitude, assuming matched filtering, of a single burst event,
for black hole masses ranging from 14M$_\odot$ (far left) to
4M$_\odot$ (far right) in steps of 1M$_\odot$, for negative
$\dot{f}$. The source distance is 100 Mpc ($z\approx 0.02$). {\bf
Middle curve}---a model for the detector noise, expressed as a
dimensionless root-mean-square amplitude, for an advanced LIGO
detector \citep{flan98}. {\bf Bottom curve}---the characteristic
dimensionless amplitude of a single cycle for the same sources.}
\label{fig3}
\end{figure*}

\begin{figure*}
\plotone{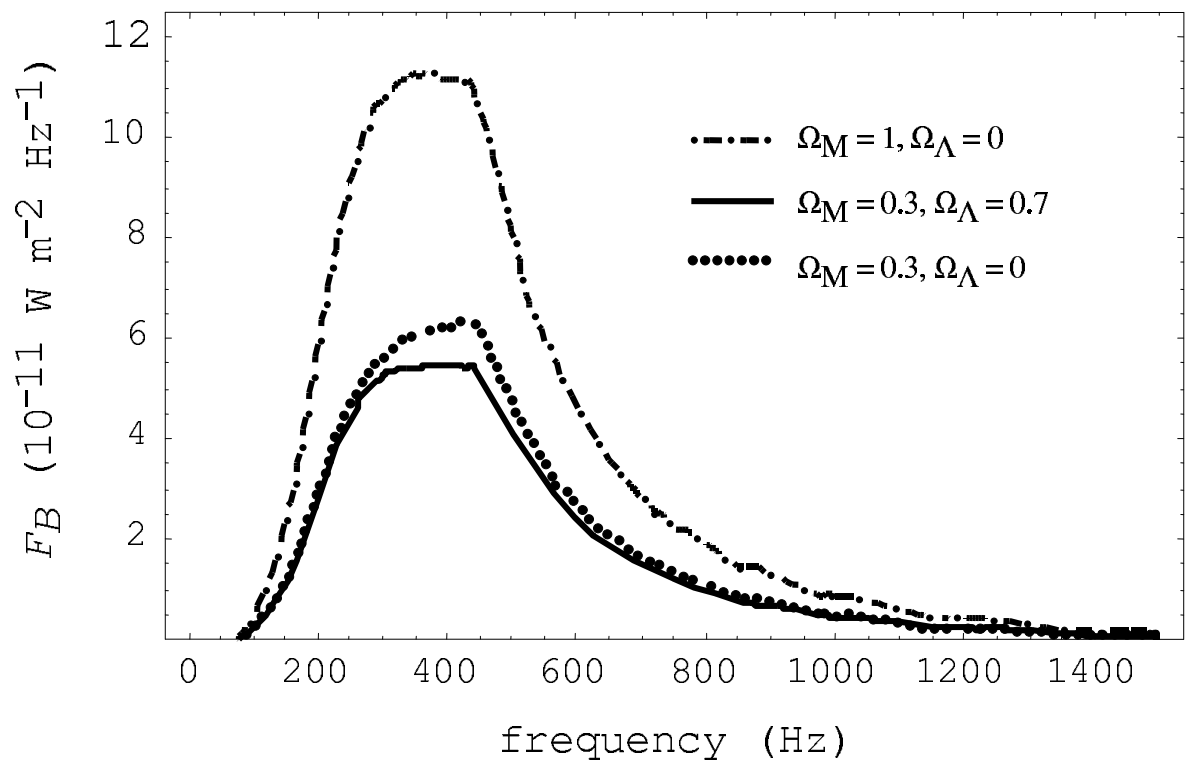}%
\caption{The stochastic background of gravitational radiation from
black hole-torus systems in the EdS, flat-$\Lambda$ and open
cosmologies, assuming a population of (4--14)M$_{\odot}$ black
holes uniformly distributed in mass and locked to the star
formation rate. For the  flat-$\Lambda$ and EdS cosmologies the
spectral flux density has a plateau, from about 300-450
Hz$\times\kappa$ with $\kappa$ a factor of order unity 
due to an uncertainty in the radius of the torus.}\label{fig4}
\end{figure*}

\begin{figure*}
\plotone{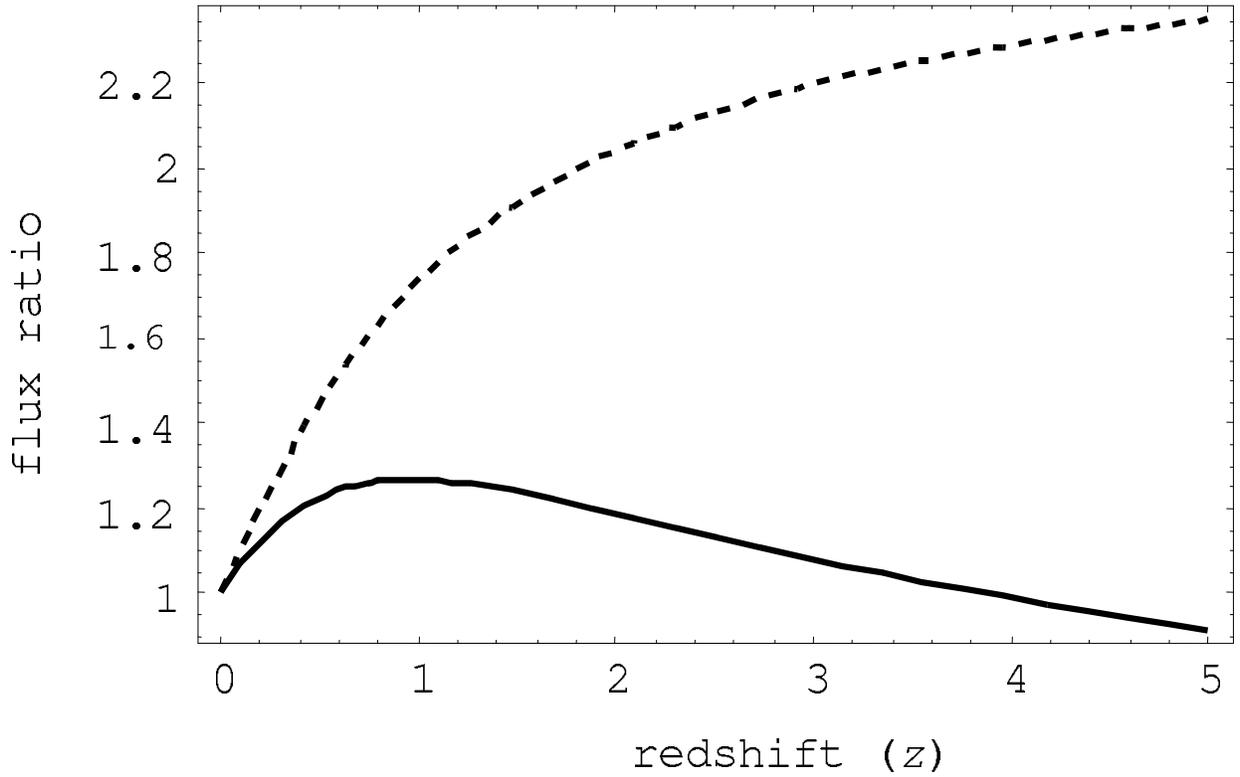}%
\caption{ The ratio of inverse luminosity distance
$d_{L}(\Omega_{\mathrm{M}},\Omega_{\Lambda})$ squared for open and
EdS to flat-$\Lambda$ cosmologies. {\bf Top
curve}---$\left[d_{L}(1,0)/ d_{L}(0.3,0.7)\right]^{-2}$. {\bf
Bottom curve}---$\left[d_{L}(0.3,0)/
d_{L}(0.3,0.7)\right]^{-2}$.}\label{fig5}
\end{figure*}

\begin{figure*}
\plotone{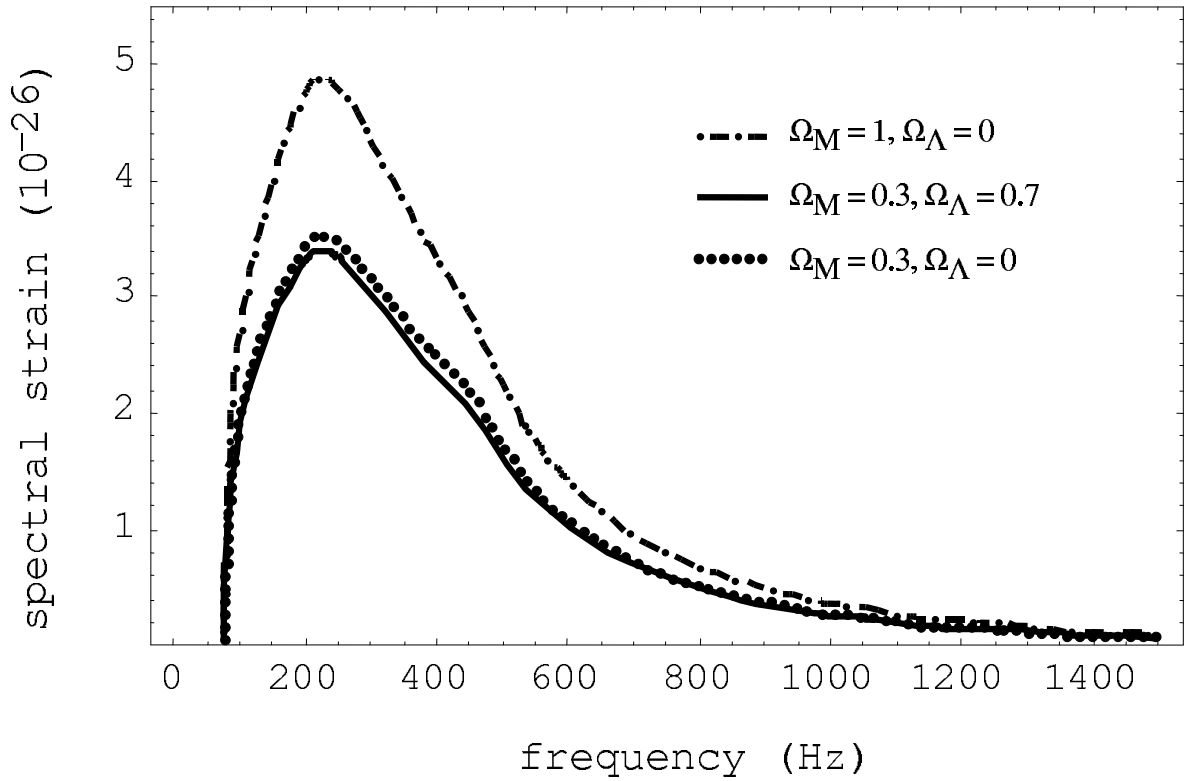}%
\caption{The background spectral strain corresponding to the
spectral flux density shown in Fig. 4, for the same cosmologies,
showing a maximum of about (3--5$)\times 10^{-26}$ Hz$^{-1/2}$ at
about 200--250 Hz.} \label{fig6}
\end{figure*}

\begin{figure*}
\plotone{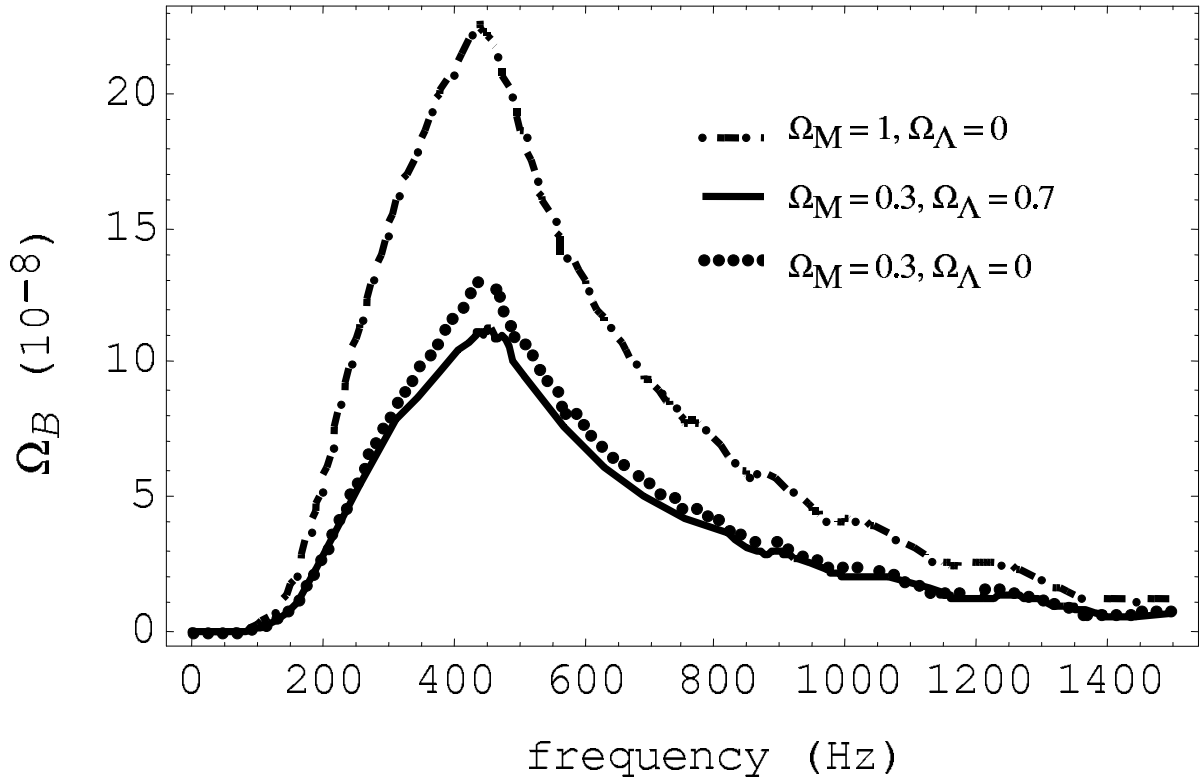}%
\caption{The spectral closure density for the background spectral
flux density shown in Fig. 4 exhibits a sharp maximum at about 450
Hz for the same cosmologies.} \label{fig7}
\end{figure*}

\begin{figure*}
\plotone{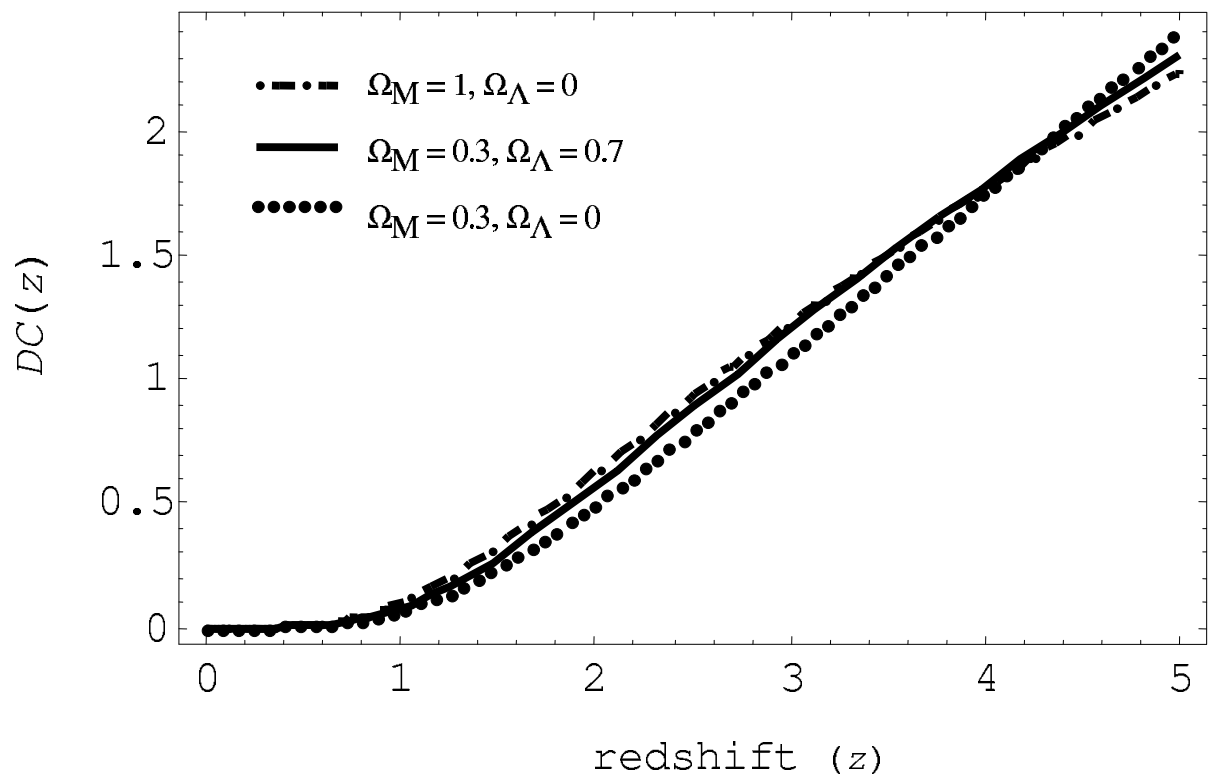} \caption{The duty cycle as a function of
redshift assuming the differential GRB rates from Fig. 1, a
typical de-redshifted duration of GW emission for each source of
20 s and a beaming factor of 500. The proposed signal is
essentially continuous, with a cumulative duty cycle of about 2
for sources out to $z=5$.} \label{fig8}
\end{figure*}

\begin{figure*}
\plotone{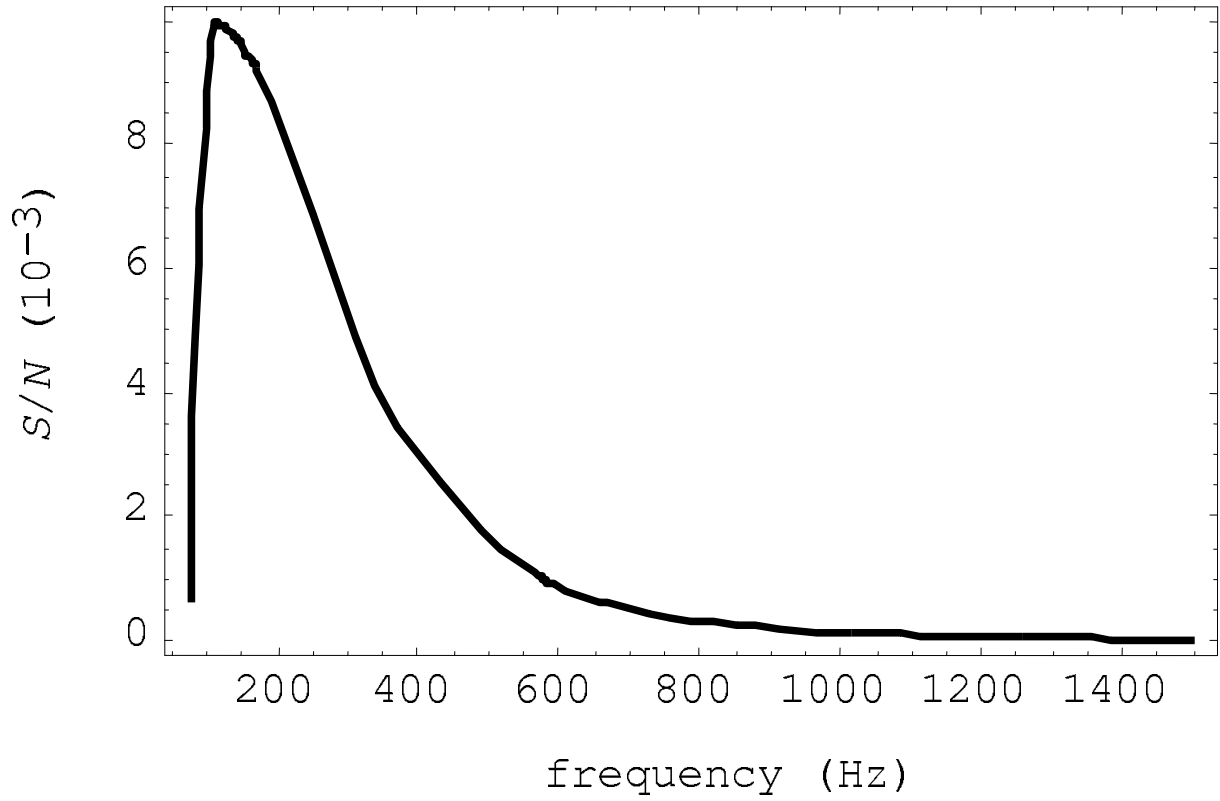} \caption{The signal-to-noise ratio for a single
interferometric detector assuming a noise power spectral density
of an advanced LIGO detector and the flat-$\Lambda$ GRB GW
background spectral closure density shown in Fig. 7. The $S/N$ is
small, the maximum being $1\times10^{-2}$ near 100 Hz.}
\label{fig9}
\end{figure*}

\begin{figure*}
\plotone{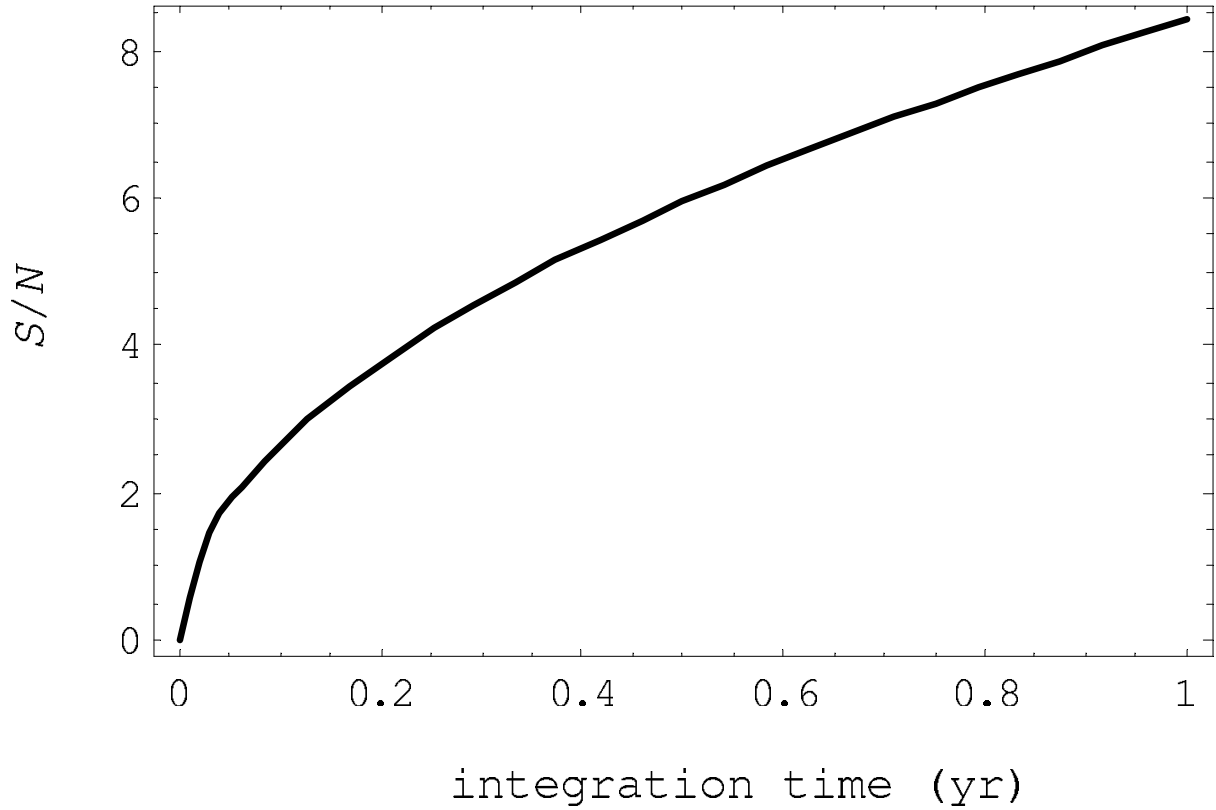} \caption{The signal-to-noise ratio as a
function of integration time for the flat-$\Lambda$ GRB background
assuming an optimized configuration of two advanced LIGO-type
detectors. Possible detectability within a year of integration is
indicated.} \label{fig10}
\end{figure*}
\end{document}